\documentclass[12pt,preprintnumber]{revtex4}
\usepackage{epsfig}
\usepackage{graphicx}
\usepackage{dcolumn}
\usepackage{bm}
\usepackage[hang,bf,small]{caption}
\usepackage{amsmath}
\DeclareGraphicsExtensions{.eps.gz,.eps,.ps,.ps.gz}
\parskip=\itemsep               

\newcommand{\beq}{\begin{equation}}
\newcommand{\eeq}{\end{equation}}
\newcommand{\be}{\begin{eqnarray}}
\newcommand{\ee}{\end{eqnarray}}

\newcommand{\as}{\alpha_S}
\newcommand{\Lb}{\left(}
\newcommand{\Rb}{\right)}

\def\eq#1{{Eq.~(\ref{#1})}}
\def\fig#1{{Fig.~\ref{#1}}}

\newcommand{\bas}{\bar{\alpha}_S}
\relax
\begin{document}
\begin{flushright}
{\tt  TAUP-2802-05} \\
{\tt \today}\\
\end{flushright}

\setcounter{footnote}{1}

\title{
{\LARGE \bf Saturation effects at LHC energies  }\\[4.5ex]
}

\author{E.~Gotsman}  \email[]{gotsman@post.tau.ac.il}
\author{E.~Levin}    \email[]{leving@post.tau.ac.il} \email[]{levin@mail.desy.de}
\author{U.~Maor}     \email[]{maor@post.tau.ac.il}
\author{E.~Naftali}  \email[]{erann@post.tau.ac.il}

\affiliation{
{\it HEP Department}\\
{\it School of Physics and Astronomy}\\
{\it Raymond and Beverly Sackler Faculty of Exact Science}\\
{\it Tel Aviv University, Tel Aviv, 69978, ISRAEL}\\
}

\centerline{}

\begin{abstract}

 Within the framework of a modified Balitsky-Kovchegov equation, we calculate and provide estimates of 
non-linear saturation effects expected in the LHC range of energies.
 \end{abstract} 
\maketitle
 \thispagestyle{empty}
 \setcounter{page}{1}


The main objective of this letter is to provide   reliable estimates for high 
density 
QCD (saturation)  effects at  the LHC range of energies.
  Our estimates are based on the 
Balitsky-Kovchegov \cite{BK} equation  which is the mean field approximation to high density QCD 
\cite{GLR,MUQI,MV}. In spite of the  restricted theoretical accuracy of this equation 
\cite{JIMWLK,IM,KOLE,LL,MUSH,IMM,LLB,IT1,MSW,LLTR,IT2,KOLU,BIT} in the framework of the 
colour dipole approach \cite{MUCD},  the physics underlying it is 
rather transparent,  and it  has led to 
a successful description of the main features of the experimental data, both for deep inelastic 
scattering and ion-ion collisions \cite{GW,KLMN,MCLIGS,GLLM,IIM}. Among the advantages of the 
Balitsky-Kovchegov equation is  the theoretical understanding of its mathematical structure 
\cite{GLR,BALE,MUPE},  the analytical solutions in the restricted domains  \cite{LT,MUTR,MUPE}, 
as well as 
extensive   experience in  numerical solutions \cite{AB,GMS,LGLM,RUWE,GOST,GKLMN,KUST, 
AAMSW,IKMCK}.
This equation is  at the moment the best that we have.

However,  the Balitsky-Kovchegov equation has  very serious deficiencies: it does not 
include  the next-to-leading corrections  to the BFKL equation \cite{BFKL} which are 
large
\cite{NLL},  and those   should  be taken into account to obtain  reliable predictions at higher 
energies 
\cite{NLLG}.  In the  estimates given here  we rely on the modified version of the Balitsky-Kovchegov 
equation
which was  suggested in Ref. \cite{GLMMOD}.  The modified Balitsky-Kovchegov equation
includes the correct  re-summation of the NLO corrections to the BFKL equation suggested by the Florence 
group \cite{FG}. It also contains   the important  observation of the Durham 
group\cite{DG},  that the main 
part 
of the 
NLO correction is  the inclusion of the leading order anomalous dimension of the DGLAP 
equation \cite{DGLAP}, this  in the framework of the BFKL approach.

 The  modified Balitsky-Kovchegov equation suggested\footnote{We have solved a slightly different 
equation (see Eq.(3.1) in  Ref. \cite{GLMMOD}) The equation, that has been solved, is a simplified 
version of \eq{MODBK}  due to numerical difficulties.} is
\beq \label{MODBK}
\frac{\partial N\Lb r,Y;b \Rb}{\partial\,Y}\,\,\,=\,\,
 \frac{C_F\,\as}{\pi^2}\,\,\int\,\frac{d^2 r'\,r^2}{(\vec{r}
\,-\,\vec{r}\:')^2\,r'^2} \,\Lb \,1\,\,-\,\,\frac{\partial}{\partial\,Y} \Rb
\left[ \,2  N\Lb r',Y;\vec{b} \, -\,
\frac{1}{2}\,(\vec{r} - \vec{r}\:')\Rb\,\,- \right.
\eeq
$$
\left. -\,
\,\, N\Lb r,Y;\vec{b} \Rb \,\,-\,\,
 N\Lb r',Y;\vec{b} - \frac{1}{2}\,(\vec{r} - \vec{r}\:')
\Rb\, N\Lb \vec{r} -
\vec{r}\:',Y;b - \frac{1}{2} \vec{r}\:'\Rb  \,
\right]
$$

\eq{MODBK} includes the full anomalous dimension of the DGLAP equation in  leading order,  in the
approximate  
form proposed in Ref. \cite{EKL}, namely,
\beq \label{GAMBFKL}
\gamma (\omega) \,\,=\,\,\bas\,\Lb \,\frac{1}{\omega} \,-\,1 \Rb
\eeq
which describes the exact $\gamma (\omega)$ within an accuracy $\leq\,5\%$. The advantage of 
\eq{MODBK} is 
 that it conserves  energy even when including  the non-linear term.

\centerline{\bf Unintegrated structure function:}
Solving \eq{MODBK} we obtain the dipole scattering amplitude.
 The more transparent physical meaning 
is the 
so called unintegrated structure function, which is  the probability for a hadron to have a gluon 
with fixed
transverse momentum.  This function enters the calculation of the
 inclusive production for the
 gluon jet,  
as well as all other inclusive cross sections.  \cite{GLR,INCL,BRUN},
\beq \label{INCXS}
\frac{d^2 \sigma}{d y\,d^2 p_t}\,\,\,=\,\,\frac{4\, \,N_c\,\as}{N^2_c - 
1}\,\frac{1}{p_t^2}\,\int\,\phi(Y-y,k_t)\,
\phi(y, \vec{p}_t - \vec{k}_t)d^2k_{t}
 \eeq 
where the unintegrated structure function $\phi$ is determined from the solution of \eq{MODBK} by the following 
equation \cite{BRUN}
\beq \label{UNIN}
\phi(Y,k_t)\,\,\,=\,\,\frac{N_c}{4\,\as(k^2_t)\,\pi^2}\,\,k^2_t \, \int\,
d^2 b\,\,\Delta_{k_t}\, N(Y; 
k_t,b)
\eeq
where $\Delta_k$ is a two dimensional Laplacian and 
\beq \label{FTR}
N(Y;k_t,b)\,\,=\,\,\int\,\frac{ d^2 r}{(2\,\pi)^2 \,r^2}\,\,e^{i \vec{k}_t \cdot 
\vec{r}}\,\,\,N(Y;r,b)\,.
\eeq
and  the gluon structure function is equal to
\beq \label{XG}
\as(Q^2)\,xG(x,Q^2)\,\,=\,\,\int^{Q^2}\,d \,k^2_t\,\,\as(k^2_t)\,\,\phi(Y,k_t)
\eeq

\begin{figure}[ht]
    \begin{center}
        \includegraphics[width=0.60\textwidth,angle=270]{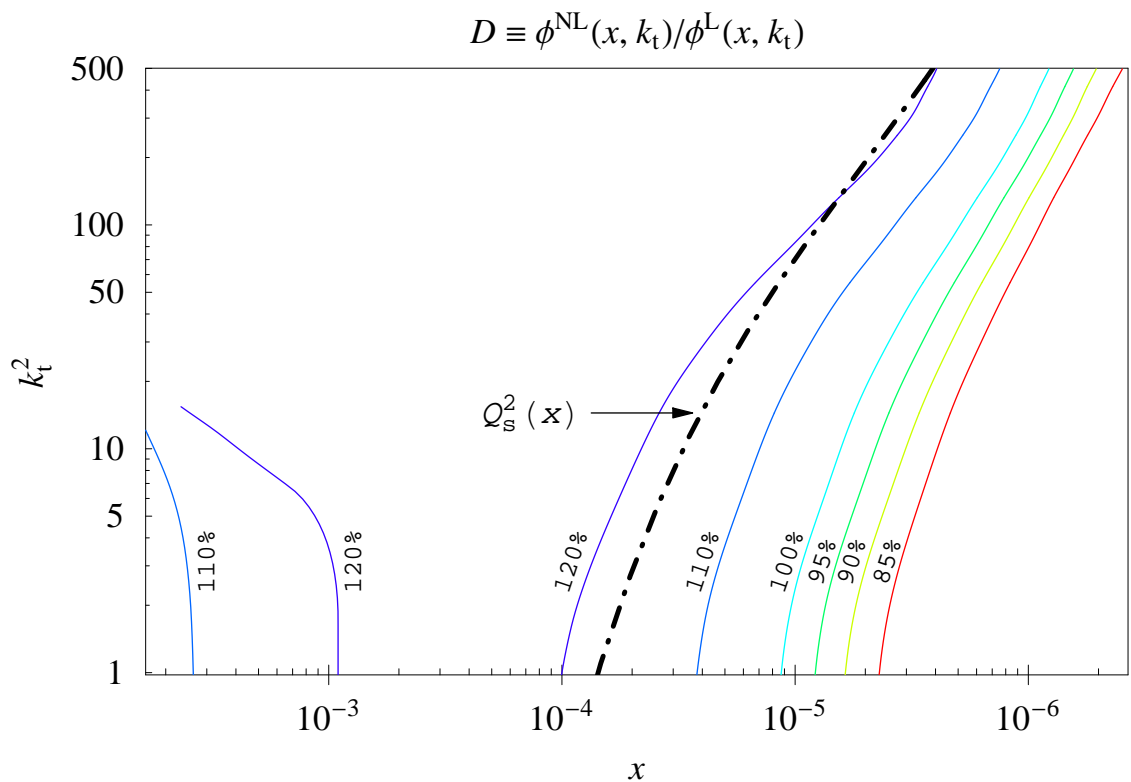}
        \caption{\it The ratio given by \eq{R} which shows the effect
          of the non-linear terms on the prediction for the dipole
          scattering amplitude at the LHC energies.} 
\label{sup}
    \end{center}
\end{figure}

\newpage
\centerline{\bf Suppression due to the saturation effects:}  
In \fig{sup} we plot the ratio
\beq \label{R}
D\Lb k_t;x \Rb\,\,=\,\,\frac{\phi^{NL} \Lb k_t, x \Rb}{\phi^{L} \Lb k_t,
x \Rb}
\eeq
where $\phi^{NL}$ is the unintegrated structure function  obtained using
all terms of Eq.(1)
,  while 
$\phi^{L}$ is, the unintegrated structure function obtained by
  excluding  the 
non-linear part of Eq.(1). 

Our initial condition for solving \eq{MODBK}, are the dipole-proton 
amplitude from the CTEQ parameterization for the gluon structure
function.  We use the following form  
of this initial distribution 
\beq \label{INCON}
N\Lb r,x;b  \Rb\,\,=\,\,\,1\,\,-\,\,exp\Lb\, - \frac{\as 
\pi^2}{6}\,r^2\,xG^{CTEQ(6)}(x,4/r^2)\,S(\sqrt{r^2+b^2})/S(0)\Rb
\eeq
where $S(b)$ is a dipole-like profile,  $R$ is the size of the proton,
we took $\,R^2 \,=\,10\,GeV^{-2}$.  

From \fig{sup} one can see that the non-linear term in \eq{MODBK} is
important, and in the LHC range of energy it suppresses the value of
$\phi$ by 30\%. It should be stressed that the inclusive cross section
is proportional to $\phi^2$, and we expect even a larger suppression
for the inclusive production (approximately twice as large).  Note
that $D(x,k_t) < 1$ for $k^2_t< Q^2_s$.  For $k_t^2\approx Q_s^2$, on
the other hand, we observe an antishadowing effect on which we shall
comment below
\footnote{We define the saturation momentum as the value of $Q=2/r$
at which $2\log(1-N(Y,r;b=0))=-1$.  In order to compare $Q_s^2$ with
the ratio $D(x,k_t)$, which is an integrated quantity, we have
normalized $Q_s^2$ such that at the leftmost point, it coincides with
the saturation momentum of the GBW model \cite{GW}.}.
Such behaviour of the ratio was expected theoretically (see for
example Refs.  \cite{GW,KLMN,MCLIGS,MUTR}), however there are no
reliable estimates in the approach which is based on the solution of
the Balitsky-Kovchegov equation, and which describes all data at lower
energies.

\begin{figure}[ht]
    \begin{center}
        \includegraphics[width=0.60\textwidth, angle=270]{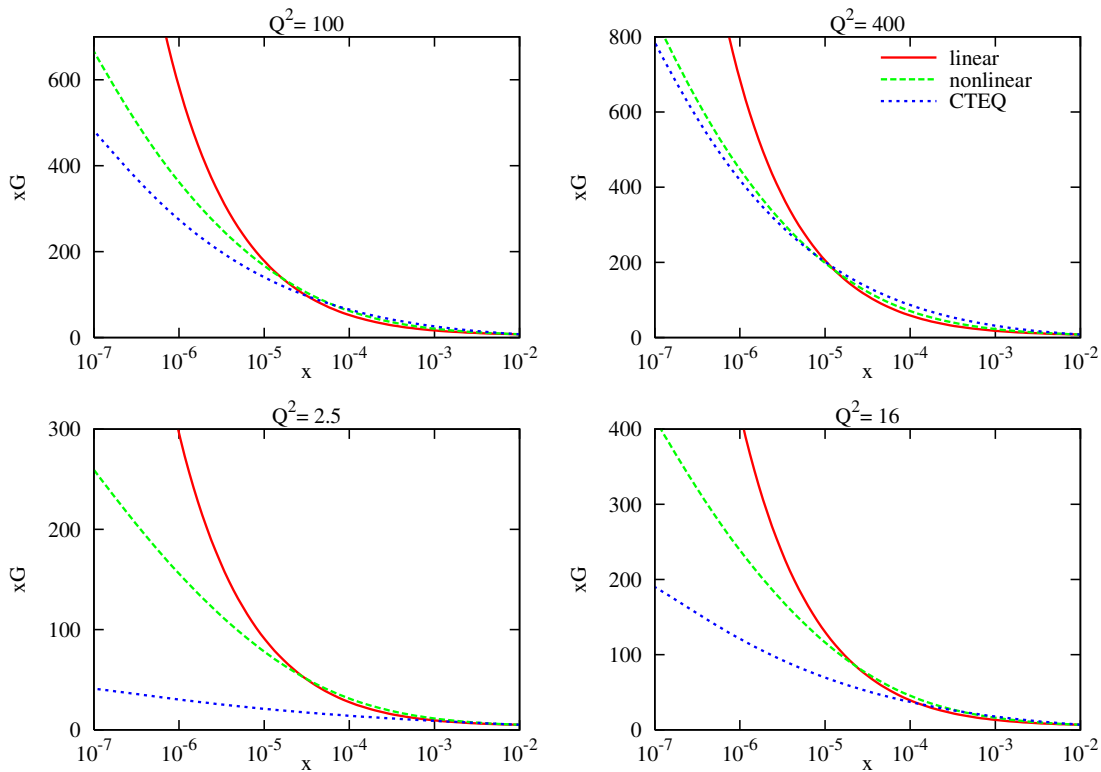}
        \caption{\it The value of the gluon  structure function  as function
 of $x$ for 
different values of $Q^2$. Non-linear  denotes the curve for the
gluon structure function 
calculated using \eq{XG} from  the solution of \eq{MODBK},  while 
linear  is the same without the non-linear term. CTEQ  denotes the
 curve for the gluon  structure function given by CTEQ parameterization.} 
\label{N}
    \end{center}
\end{figure}

\begin{figure}[ht]
    \begin{center}
        \includegraphics[width=0.60\textwidth, angle=270]{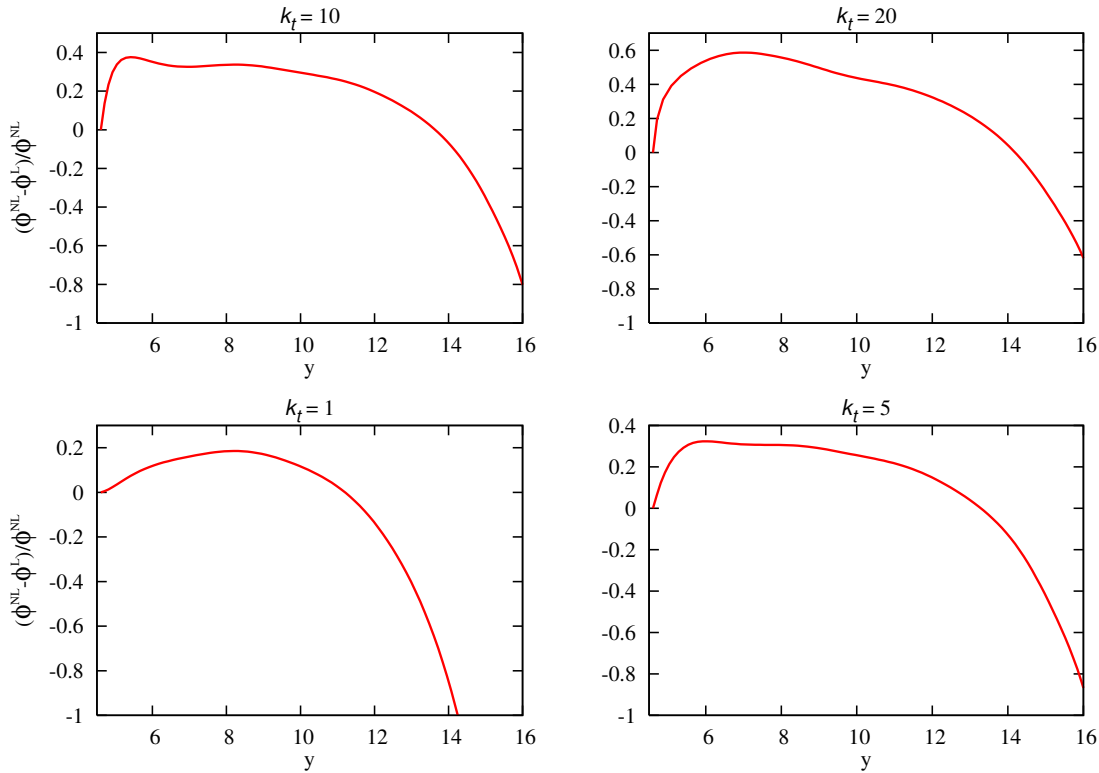}
        \caption{\it  The ratio $(\phi^{NL} - \phi^{L})/\phi^{NL}$  for different values of $Q^2$ as 
function of $y = \ln(1/x)$.}
\label{antshad}
    \end{center}
\end{figure}

\centerline{\bf Increase  due to the saturation effects (antishadowing):} 
\fig{N} shows the value of $xG(x,Q^2)$ for the non-linear equation (see
\eq{MODBK}),  together with the value for the solution of the linear equation
(Eq.(1) without the non-linear terms).  Note that the solution to the
non-linear equation leads to a larger amplitude for large values of $x$.  This
is not  expected as the non-linear term has a negative sign, and manifests
itself as the shadowing correction which  decreases  the value of the
amplitude. The increase of the ratio $D$ (see \eq{R} and  \fig{antshad}) at
low energies is significant, and cannot be explained by  errors due to the
approximate formula  for the anomalous dimension of the DGLAP equation, or due
to the numerical prodecure.

The explanation for the anti-shadowing effect is very simple. At large
$k_t$, the non-linear term is small and the linear evolution equation
is a good approximation to the solution. Decreasing the value of
$k_t$, the largest changes are induced in the slope of the amplitude
(or/and gluon densities) with respect to $\ln (k_t)$ (see
Refs. \cite{GLR,MUTR,MCLIGS,KLMN}). The slope for the linear equation
is larger than for the non-linear equation. 
For large values of $k_t$, the unintegrated structure function $\phi$ is equal to the solution of 
the linear equation, while for small values of $k_t$, $\phi$ becomes larger than the solution to the 
linear equation.

Another way to see the same effect is by  comparing derivatives with
respect to Y see \fig{N}. For the solution to the non-linear equation at large $x$ the
derivatives with respect to $Y$ are much larger than the same derivative for the solution of the 
non-linear equation. Therefore,\,    for the case of the non-linear equation  the linear term, which 
contains $- \partial/\partial Y$, is larger for the non-linear equation, leading
to an increase in the value of the solution.   The unintegrated structure function in \eq{UNIN}
reproduces the same qualitative behaviour as the dipole scattering
amplitude.

All these arguments can be obtained directly from energy-momentum sum rules which 
hold for  the modified Balitsky-Kovchegov equation, namely, that the integral
over $x$ is independent of  $Q^2$ 
\beq \label{ESR}
\int\,\,d\,x \phi^{NL}\Lb Y;Q^2 \Rb\,\,=\,\,Const(Q^2)\,\,=\,\,\int\,\,d\,x \phi^{L}\Lb Y;Q^2 \Rb
\eeq

In the region $k^2_s \,<\,Q^2_s(x)$ (or $x < x_s$,  where $ x_s $ is the
solution of the equation $k^2_t = Q^2_s(x_s)$), we expect $D(x,k_t) < 1 $ (see
\fig{N}) and based on \eq{R},  we anticipate that in some region of $x > x_s$,
this ratio   should be larger than 1. 

\fig{antshad} displays  quantitative estimates for these antishadowing effects.

The effect is rather large and has been discussed previously  in Ref.
 \cite{ANTSHAD}.   In 
our approach it  arises
naturally \footnote{In the talk 
of K. Peters at HERA-LHC Workshop (19  January 2005), it was claimed that there is no  antishadowing 
effect.  As far as we understood from his transparencies,  he used the modification to the 
Balitsky-Kovchegov equation suggested in Ref. \cite{KU}. This modification is quite different from 
ours and does not preserves energy conservation in both the  linear and non-linear terms. This is 
the reason why no  anti-shadowing effect was  seen.}.

The comparison of the results of the present calculation with those of CTEQ and MRST 
 is not straight forward, since we calculated the unintegrated gluon distribution, 
while CTEQ and MRST present results only for the structure functions. We are presently busy with 
this project and hope to be able to present our results in the near future.

We hope that our estimates of the scale for the nonlinear effects at the LHC energies, will help 
to produce  more reliable calculations  for the cross sections of `hard' processes of interest at high 
energies.

{\bf Acknowledgments:}
We  would like to thank Jochen Bartels, Markus Diehl, Hannes Jung, Michael Lublinsky   and all 
participants of  HERA-LHC 
workshop   for fruitful
 discussions.  This research was supported in part  by the Israel Science Foundation, founded by the
 Israeli Academy of Science and Humanities.

\end{document}